\documentclass[showkeys,showpacs,preprintnumbers,aps,twocolumn]{revtex4}
\usepackage{epsfig}

\begin{document}

\date{11.12.2001}

\title{Chaotic front dynamics in semiconductor superlattices}

\author{A. Amann}
\author{J. Schlesner}
\author{A. Wacker}
\author{E. Sch{\"o}ll}
\affiliation{Institut f{\"u}r Theoretische
Physik, Technische Universit{\"a}t Berlin,
Hardenbergstrasse 36, 10623, Berlin, Germany}

\begin{abstract}
We analyze the dynamical evolution of the current and the charge density in a 
superlattice for fixed external dc voltage in the regime of self-sustained 
current oscillations, using a microscopic sequential tunneling model.
Fronts of accumulation and depletion
layers which are generated at the emitter may collide and annihilate,
thereby leading to a variety of different scenarios. We
find complex chaotic regimes at high voltages and low contact conductivities.
\end{abstract}


\maketitle

\section{Introduction}
\label{intro}

Electronic transport in semiconductor superlattices (SL) is known to 
show strongly nonlinear spatio-temporal dynamics. Either self--sustained 
current oscillations \cite{KAS95,HOF96,KAS97,PAT98,STE00a} or a
sawtooth--like current--voltage characteristic with many branches
associated with static field domains \cite{ESA74,KAW86,GRA91,SCH00} 
have been found. For a recent review see \cite{WAC01}.
Under time-dependent external voltage conditions, superlattices exhibit
a rich menagerie of complex behavior including ac driven chaos and 
switching scenarios between multistable states. This was 
studied recently experimentally \cite{ZHA96,KAS96,LUO98,SHI97a,ROG01} and 
theoretically \cite{BUL95,AMA01}. 

In this paper we present simulations of dynamic scenarios for superlattices
under fixed time-independent external voltage in the regime where 
self-sustained dipole
waves are spontaneously generated at the emitter. The dipole waves are 
associated with traveling field domains, and
consist of electron accumulation and depletion fronts which in general
travel at different velocities and may merge and annihilate.
We find that depending on the applied
voltage and the contact conductivity, this gives rise to various oscillation 
modes and self-synchronization effects as well as different routes to chaotic
behavior.

A similar scenario of merging fronts was recently found in the context of a 
spatially continuous model describing bulk impurity impact ionization breakdown
\cite{CAN01}. It is also reminiscent of patterns of temperature 
pulses in globally coupled heterogeneous catalytic systems, e.g. 
\cite{GRA93a}.  

\section{The Model}
\label{sec:model}

\begin{figure}[htbp]
  \begin{center}
    \epsfig{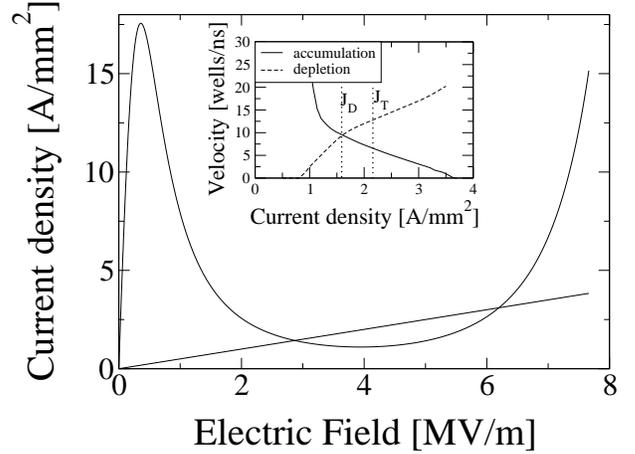}
    \caption{Current density vs electric field
      characteristic at the emitter barrier (straight line) and
      between two neutral wells. The Ohmic conductivity of the emitter
      is $\sigma = 0.5 \quad\Omega^{-1} m^{-1}$.  The inset shows the front 
      velocity vs.\ current density for electron depletion and accumulation
    fronts. $J_D$ and $J_T$ denote the current for dipole and tripole
    propagation, respectively.}    
    \label{fig:homogen_velocity}
  \end{center}
\end{figure}

Weakly coupled superlattices are successfully described by a one
dimensional sequential tunneling model for electrons
\cite{PRE94,BON94,KAS97}.
In the framework of this model electrons are assumed to be localized
at one particular well and only weakly coupled to the neighboring
wells. The tunneling rate to the next well is lower than the typical
relaxation rate  between the different energy levels within one well.
The electrons within one well are then in quasi--equilibrium and
transport through the barrier is incoherent. The resulting
tunneling current density $J_{m\to m+1}(F_m, n_m, n_{m+1})$ from well
$m$ to well $m+1$ depends only on the electric field $F_m$ between
both wells and the electron densities $n_m$ and $n_{m+1}$ in the wells
(in units of $cm^{-2}$). For
details of the microscopic calculation of $J_{m\to m+1}$ we refer 
to the literature \cite{AMA01,WAC01}. A typical result
for the current density vs electric field characteristic is depicted in
Fig.~\ref{fig:homogen_velocity} in the spatially homogeneous case,
i.e. $n_m=n_{m+1}=N_D$, with donor density $N_D$.

In the following we will adopt the total number of electrons in one well as
the dynamic variables of the system. The dynamic equations are then
given by the continuity equation
\begin{equation}
  \label{eq:continuity}
  e\frac{\text{d}n_m}{\text{d}t} = J_{m-1 \to m}  - J_{m \to m+1}
\quad \text{for } m  = 1, \ldots N
\end{equation}
where $N$ is the number of wells in the superlattice. 

The electron densities and the electric fields are coupled by
the following discrete version of Gauss's law
\begin{equation}
  \label{eq:poisson}
  \epsilon_r \epsilon_0 (F_{m} - F_{m-1}) = e(n_m -N_D)
\quad \text{for } m  = 1, \ldots N,
\end{equation}
where $\epsilon_r$ and  $\epsilon_0$ are the relative and absolute 
permittivities, $e<0$ is the electron charge, and
$F_0$ and $F_N$ are the fields at the emitter and collector
barrier, respectively.

The applied voltage between emitter and collector gives rise to a
global constraint
\begin{equation}
  \label{eq:voltage}
  U =  \sum_{m=0}^N F_m d,
\end{equation}
where $d$ is the superlattice period.

The current densities at the contacts are chosen such that dipole waves are 
generated at the emitter. For this purpose it is sufficient to choose Ohmic
boundary conditions:
\begin{eqnarray}
  \label{eq:boundary}
  J_{0 \to 1} &=& \sigma F_0 \\
  J_{N \to N + 1} &=& \sigma F_N \frac{n_N}{N_D}
\end{eqnarray}
where $\sigma$ is the Ohmic contact conductivity, and the factor $n_N/N_D$ is
introduced in order to avoid negative electron densities at the
collector. Here we make the physical assumption that the current from
the last well to the collector is proportional to the electron density
in the last well. It is in principle possible to calculate the boundary
conditions using microscopic considerations \cite{AGU97,BON00}, but the 
qualitative behavior is not changed.

\section{Numerical results}
\label{sec:numeric}

\begin{figure}[htbp]
  \begin{center}
    \epsfig{file=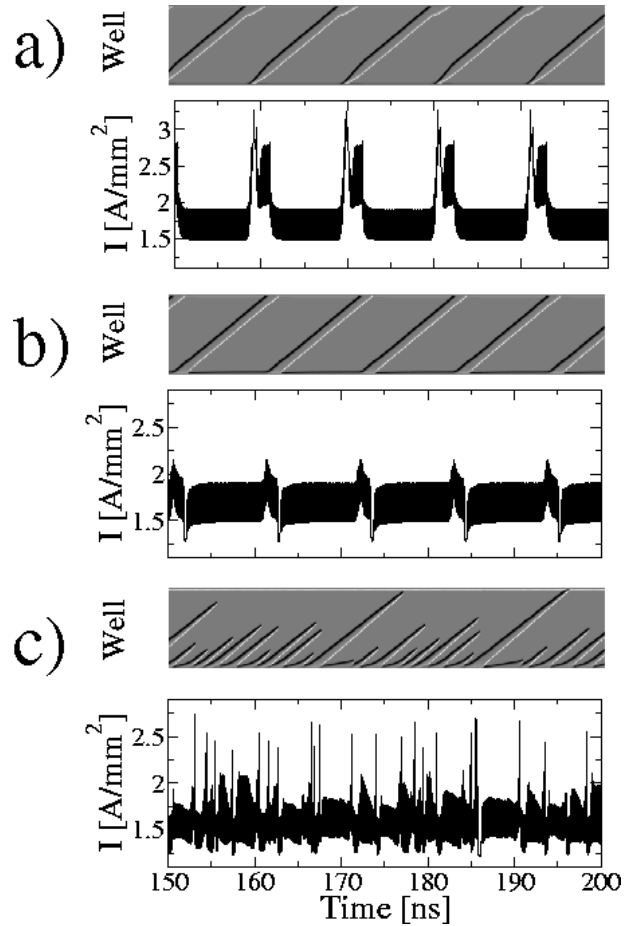,width=0.45\textwidth}
    \caption{Dynamic evolution of charge density and current $I$ for 
      contact conductivities (a) $\sigma=1.3 \Omega^{-1} \text{m}^{-1}$,
      (b) $\sigma=0.6 \Omega^{-1} \text{m}^{-1}$ and 
      (c) $\sigma=0.55 \Omega^{-1} \text{m}^{-1}$ for fixed bias $U=1.0\text{V}$.
      Light and dark regions denote electron
      accumulation and depletion 
      fronts in the space-time plots of the 
      charge densities, respectively.}
    \label{fig:sigma}
  \end{center}
\end{figure}

In our computer simulations we use an $N = 100$ superlattice with 
Al$_{0.3}$Ga$_{0.7}$As barriers of width $b=5 \text{nm}$ and GaAs quantum
wells of width $w=8 \text{nm}$, doping density 
$N_D= 1.0 \times 10^{11} \text{cm}^{-2}$ and scattering
induced broadening $\Gamma = 8 \text{meV}$ at $T=20 \text{K}$.  The
contact conductivity $\sigma$ is choosen such that the intersection point
with the homogeneous current density vs field characteristic in
Fig.~\ref{fig:homogen_velocity} is at a current value at which no
stationary field domain boundaries exists. By this configuration, accumulation
and depletion fronts are generated at the emitter.  For large values of $\sigma$,
e.g. $\sigma =
1.3 \quad \Omega^{-1} \text{m}^{-1}$, we find that those fronts form a
dipole, i.e. a traveling field domain, with leading electron depletion front 
and trailing accumulation
front (cf Fig.~\ref{fig:sigma}~(a)). The dipole traverses the sample at
almost constant velocity and constant current which shows up as a plateau
in the time trace of the current. The current $J_D$ and the front velocity in 
this dipole 
propagation mode are given by the intersection point of the velocities of
accumulation and depletion front (cf. inset of
Fig.~\ref{fig:homogen_velocity}). As the leading depletion front
reaches the collector, there is no speed constraint on the remaining
accumulation front. As it accelerates, the current rises, and a new
dipole is generated at the emitter. Now for a short time there are two
accumulation and one depletion fronts in the sample. In order to
fulfill eq.(\ref{eq:voltage}) for fixed $U$ the depletion front has to assume 
twice
the velocity of the accumulation fronts. This constraint fixes the
current to $J_T$ (cf. inset of Fig.~\ref{fig:homogen_velocity}) during
this tripole regime. Note that in the current time trace the fast small-amplitude 
oscillations
(due to well-to-well hopping of depletion and accumulation fronts in our
discrete model) in the dipole and tripole regime are not resolved temporally.

At lower contact conductivity  $\sigma  = 0.6 \quad \Omega^{-1}
\text{m}^{-1}$, we find that instead of a dipole only a depletion
front is generated at the emitter, as the old depletion front reaches
the collector (cf. Fig. \ref{fig:sigma}~(b)). Now for a short time a dipole with
leading accumulation front exists. The current is fixed again by the
constraint of equal velocities of accumulation and depletion front, 
as for the dipole with leading depletion front.
At the time the old accumulation
front reaches the collector a new accumulation front is generated at
the emitter. This process is accompanied by a dip in the current time trace.

For even lower  $\sigma  = 0.55 \quad \Omega^{-1} \text{m}^{-1}$ the fronts at
the emitter are generated as dipoles with leading accumulation 
and trailing depletion fronts  (cf. Fig. \ref{fig:sigma}~(c)). The velocity of
the fronts is 
again determined by the current and the number of fronts in the
sample. Since the two types of fronts in general move at
different velocities, merging and annihilation of accumulation and
depletion fronts may occur, which may lead to complicated behaviour
including chaos as shown in Fig.~\ref{fig:sigma}~(c). 

\begin{figure}
  \begin{center}
    \epsfig{file=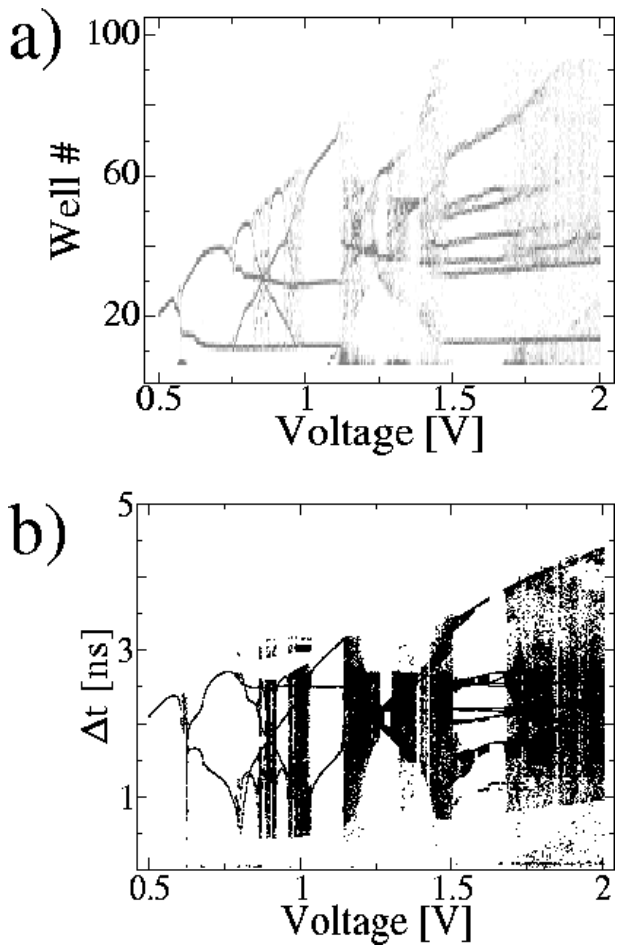,width=0.45\textwidth}
    \caption{(a) Positions where accumulation and depletion fronts
      annihilate vs voltage at $\sigma=0.5\quad \Omega^{-1} \text{m}^{-1}$.
      The grayscale indicates high (black) and low (white) numbers of 
      annihilations
      at a given well. (b) Time differences between
      consecutive maxima of the electron density in well no. 20 vs voltage at 
      $\sigma=0.5\quad \Omega^{-1}\text{m}^{-1}$. Time series of length $600 ns$ have been 
      used for each value of the voltage.
      }
    \label{fig:positions_delta_t}
  \end{center}
\end{figure}

In order to study the bifurcation scenario leading to chaos we now fix
the  boundary conductivity to $\sigma= 0.5\quad \Omega^{-1}\text{m}^{-1}$. 
Experimentally, low contact conductivities may be realized by changing the
doping and barrier structure of the emitter. The temperature dependence of 
the emitter current may also be exploited.
In Fig.~\ref{fig:positions_delta_t} (a) a density plot of the positions 
(well numbers)
at which two fronts annihilate is shown as a function of the voltage. We see 
that for low
voltage the annihilation takes place at one definite position in the superlattice 
with a variation of only a few wells. This
distribution broadens for increasing voltage in characteristic
bifurcation scenarios reminiscent of period doubling, leading to chaotic
regimes. We note that in the chaotic region periodic windows exist. 
Since we are dealing with a discrete system, the position of the
merging of two fronts is also discrete. For more accurate analyses
it is convenient to use a  continuous variable such as the time difference 
between two maxima in the
electron density in a specific well. The corresponding bifurcation
diagram for well no.~20 is shown in Fig.~\ref{fig:positions_delta_t}~(b). It exhibits 
a complex structure.
\begin{figure}
  \begin{center}
    \epsfig{file=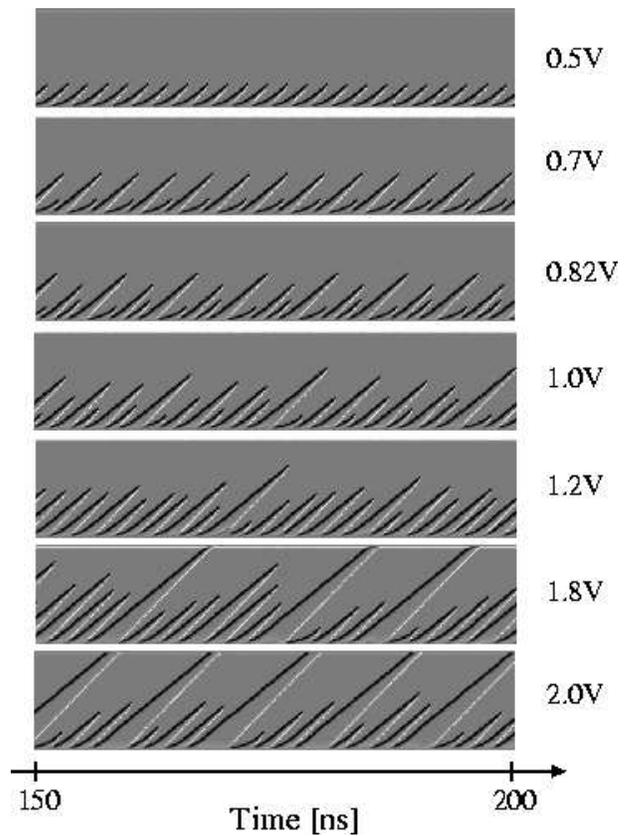,width=0.45\textwidth}
    \caption{Dynamic evolution of charge density for various
      voltages at $\sigma=0.5\quad \Omega^{-1} \text{m}^{-1}$.  Light and 
      dark regions denote electron accumulation
      and depletion layer, respectively.}
    \label{fig:density_chaos}
  \end{center}
\end{figure}

The transition from periodic to chaotic oscillations is enlightened by
considering the space-time plots for the evolution of the electron densities at 
different
voltages $U$ (Fig.~\ref{fig:density_chaos}). While for $U =
0.5 \text{V}$ the oscillations are regular, we see that at $U =0.7 \text{V}$
long and short front patterns alternate, which is characteristic for
a period doubling bifurcation. Every second pair of fronts travels farther
into the superlattice before they annihilate.
At  $U =0.79 \text{V}$ a further period
doubling occurs, as can be most clearly seen at the alternating length of the 
longer front patterns.
This yields a period-four cycle. Further increase of the voltage
finally leads to well developed chaos, with front patterns of different
lengths. For $U \geq 1.8 \text{V}$ we find that fronts may even traverse
the sample, while the dynamics remains chaotic.

\section{Summary}

We have shown that complex chaotic spatio-temporal scenarios may arise in 
weakly coupled
superlattices under {\em time-independent} external voltage conditions, 
whereas previous studies have focussed on ac-driven chaos.

\vspace{0.2cm}
This work was partially supported by DFG in the framework of Sfb 555.
We greatfully acknowledge discussion with L. L. Bonilla, M. Rogozia
and E. Schomburg.


\begin{thebibliography}{10}

\bibitem{KAS95}
J. Kastrup, R. Klann, H.~T. Grahn, K. Ploog, L.~L. Bonilla, J. Gal{\'a}n, M.
  Kindelan, M. Moscoso, and R. Merlin, Phys.~Rev.~B {\bf 52},  13761  (1995).

\bibitem{HOF96}
K. Hofbeck, J. Grenzer, E. Schomburg, A.~A. Ignatov, K.~F. Renk, D.~G.
  Pavel'ev, Y. Koschurinov, B. Melzer, S. Ivanov, S. Schaposchnikov, and P.~S.
  Kop'ev, Phys.~Lett.~A {\bf 218},  349  (1996).

\bibitem{KAS97}
J. Kastrup, R. Hey, K.~H. Ploog, H.~T. Grahn, L.~L. Bonilla, M. Kindelan, M.
  Moscoso, A. Wacker, and J. Gal{\'a}n, Phys.~Rev.~B {\bf 55},  2476  (1997).

\bibitem{PAT98}
M. Patra, G. Schwarz, and E. Sch{\"o}ll, Phys.~Rev.~B {\bf 57},  1824  (1998).

\bibitem{STE00a}
H. Steuer, A. Wacker, E. Sch{\"o}ll, M. Ellmauer, E. Schomburg, and K.~F. Renk,
  Appl.~Phys.~Lett. {\bf 76},  2059  (2000).

\bibitem{ESA74}
L. Esaki and L.~L. Chang, Phys.~Rev.~Lett. {\bf 33},  495  (1974).

\bibitem{KAW86}
Y. Kawamura, K. Wakita, H. Asahi, and K. Kurumada, Jpn.~J.~Appl.~Phys. {\bf
  25},  L928  (1986).

\bibitem{GRA91}
H.~T. Grahn, R.~J. Haug, W. M{\"u}ller, and K. Ploog, Phys.~Rev.~Lett. {\bf
  67},  1618  (1991).

\bibitem{SCH00}
E. Sch{\"o}ll, {\em Nonlinear spatio-temporal dynamics and chaos in
  semiconductors} (Cambridge University Press, Cambridge, 2001).

\bibitem{WAC01}
A. Wacker, Phys.~Rep. {\bf 357},  1  (2002).

\bibitem{ZHA96}
Y. Zhang, J. Kastrup, R. Klann, K. Ploog, and H.~T. Grahn, Phys.~Rev.~Lett.
  {\bf 77},  3001  (1996).

\bibitem{KAS96}
J. Kastrup, F. Prengel, H.~T. Grahn, K. Ploog, and E. Sch{\"o}ll, Phys.~Rev.~B
  {\bf 53},  1502  (1996).

\bibitem{LUO98}
K.~J. Luo, H.~T. Grahn, and K.~H. Ploog, Phys.~Rev.~B {\bf 57},  6838  (1998).

\bibitem{SHI97a}
Y. Shimada and K. Hirakawa, Jpn.~J.~Appl.~Phys. {\bf 36},  1944  (1997).

\bibitem{ROG01}
M. Rogozia, S.~W. Teitsworth, H.~T. Grahn, and K.~H. Ploog, Phys.~Rev.~B {\bf
  64},  041398(R)  (2001).

\bibitem{BUL95}
O.~M. Bulashenko and L.~L. Bonilla, Phys.~Rev.~B {\bf 52},  7849  (1995).

\bibitem{AMA01}
A. Amann, A. Wacker, L.~L. Bonilla, and E. Sch{\"o}ll, Phys.~Rev.~E {\bf 63},
  066207  (2001).

\bibitem{CAN01}
I.~R. Cantalapiedra, M.~J. Bergmann, L.~L. Bonilla, and S.~W. Teitsworth,
  Phys.~Rev.~E {\bf 63},  056216  (2001).

\bibitem{GRA93a}
M.~D. Graham, U. Middya, and D. Luss, Phys.~Rev.~E {\bf 48},  2917  (1993).

\bibitem{PRE94}
F. Prengel, A. Wacker, and E. Sch{\"o}ll, Phys.~Rev.~B {\bf 50},  1705  (1994),
  ibid {\bf 52}, 11518 (1995).

\bibitem{BON94}
L.~L. Bonilla, J. Gal{\'a}n, J.~A. Cuesta, F.~C. Mart\'{\i}nez, and J.~M.
  Molera, Phys.~Rev.~B {\bf 50},  8644  (1994).

\bibitem{AGU97}
R. Aguado, G. Platero, M. Moscoso, and L.~L. Bonilla, Phys.~Rev.~B {\bf 55},
  16053  (1997).

\bibitem{BON00}
L.~L. Bonilla, G. Platero, and D. S{\'a}nchez, Phys.~Rev.~B {\bf 62},  2786
  (2000).

\end{thebibliography}
\end{document}